\documentclass{iopart}
\usepackage{mathptmx}
\usepackage{bm,amsfonts}
\usepackage{youngtab}
\makeatletter
\@ifundefined{pdfoutput} 
{ 
\usepackage[dvips]{graphicx,color}
}
{ 
\ifnum\pdfoutput=0\relax
\usepackage[dvips]{graphicx,color}
\fi
\ifnum\pdfoutput=1\relax
\usepackage[pdftex]{graphicx,color}
\fi
}
\makeatother
\newcommand{\beq}{\begin{equation}}
\newcommand{\eeq}{\end{equation}}
\newcommand{\beqa}{\begin{eqnarray}}
\newcommand{\eeqa}{\end{eqnarray}}
\newcommand{\nn}{\nonumber \\}

\newcommand {\np}[1]{{\mbox{\textrm{:}\,}{#1}{\,\textrm{:}}} }

\def \e {\mathrm{e}}
\def \ex {\mathrm{e}}

\def \l {\lambda}
\def \la {\langle}
\def \o {\underline{\omega}}
\def \ra {\rangle}

\def \s {\sigma}
\def \t {\tau}

\def \Dc {\mathcal D}
\def \G   {\Gamma}
\def \I {{\mathbb{I}}}

\def \P {{\mathcal P}}

\def \Sc{{\dot{S}}}
\def \Z {{\mathbb Z}}
\def \ch {\mathrm{ch}}

\def \z {\zeta}
\def \L {\underline{\Lambda}}    \def \D {\Delta}

\def \PF {\mathrm{PF}}

\def \mod {\ \mathrm{mod} \ }
\def \H {{\mathcal H}}
\def \uu {{\widehat{u(1)}}}
\begin{document}
\title[Exact modular $S$ matrix for the $\Z_k$ parafermion quantum Hall islands]{Exact modular $S$ matrix for the 
$\Z_k$ parafermion quantum Hall islands and measurement of non-Abelian anyons}
\author{Lachezar S. Georgiev}
\address{Institute for Nuclear Research and Nuclear Energy, Bulgarian Academy of Sciences, 
 Tsarigradsko Chaussee 72,  1784 Sofia, Bulgaria}
\begin{abstract}
Using the decomposition of rational conformal filed theory characters for the $\Z_k$ parafermion quantum Hall droplets for 
general $k=2,3, \ldots$, we derive analytically the full modular $S$ matrix for these states,
including the $\uu$ parts corresponding to the charged sector of the full conformal field theory and the neutral parafermion contributions 
corresponding to the diagonal affine coset models.
This precise neutral-part parafermion $S$ matrix is derived from the explicit relations between the coset matrix and those for the 
numerator and denominator of the coset and the latter is expressed in compact form due to the level--rank duality between the affine
Lie algebras $\widehat{su(k)_2}$ and $\widehat{su(2)_k}$. The exact results obtained for the $S$ 
matrix elements are expected to play an important role for identifying interference patterns of fractional quantum Hall states 
in Fabry--P\'erot interferometers which can be used to distinguish between Abelian and non-Abelian statistics of quasiparticles 
localized in the bulk of  fractional quantum Hall droplets as well as for nondestructive interference measurement of Fibonacci anyons
which can be used for universal topological quantum computation.
\end{abstract}
\pacs{71.10.Pm, 73.21.La, 73.23.Hk,  73.43.--f}
\noindent{\it Keywords\/}:  Parafermion quantum Hall states, non-Abelian anyons, non-Abelian interference

\section{Introduction}
The quasiparticle excitations in some two-dimensional fractional quantum Hall (FQH) states  are believed to obey non-Abelian 
exchange statistics, however, so far this exotic possibility has not been proven experimentally. 
An interesting hierarchy of such non-Abelian FQH states, corresponding to FQH filling factors
\beq \label{nu}
\nu_H=\frac{k}{k+2}, \quad k=1, 2, \ldots, 
\eeq
 in the second Landau level\footnote{so the total filling factor is $2+\nu$ or $3-\nu$ for the particle--hole conjugate states}, 
has been  proposed by Read and Rezayi \cite{rr}, by constructing analytically the many-body 
electron wavefunctions with a number of non-Abelian quasiparticles  as correlation functions of the two-dimensional 
conformal field theories (CFT) known as the $\Z_k$ parafermions \cite{fat-zam}.
The non-Abelian anyons, should they exist in Nature, are capable of  topologically protected quantum 
information processing \cite{sarma-freedman-nayak,sarma-RMP}. For example, the $\Z_3$ parafermion FQH state could
realize the so called Fibonacci anyons, whose braid matrices generate a dense subgroup of the unitary 
group and could therefore be used for the implementation of a universal topological quantum computer \cite{sarma-RMP}. 

In this paper we will follow the conformal filed theory approach of Ref.~\cite{bonderson-12-5} to describe the interference patterns of the 
$\Z_k$ parafermion quasiparticles in electronic Fabry--P\'erot \cite{nayak-NA-interferometer} or Mach--Zehnder interferometers 
\cite{bais-mach-zehnder,bonderson-5-2}. These interference patterns could give a convenient  experimental signature
of the presence of non-Abelian anyons in FQH islands formed by two quantum point-contacts inside of a FQH bar. The modular $S$
matrix for the $\Z_k$ parafermions have been computed previously  in  Ref.~\cite{bonderson-5-2} using the $su(2)_k/u(1)$ coset construction.
However, the diagonal coset construction of Ref.~\cite{NPB2001} is physically more intuitive and  more appropriate for the realization of 
$\Z_k$ parafermions in 
FQH states. In particular it explains the presence of the $\Z_k$ pairing rule in the $\Z_k$ parafermion FQH states as being inherited from 
their Abelian parents, where this rule emerges naturally in search of a charge--neutral degrees of freedom decomposition \cite{NPB2001}. 
In any case, for the description 
of the interference patterns it is important to know the precise modular $S$ matrix, not that matrix upto complex conjugation, 
or upto the action of some simple currents \cite{CFT-book}, which are known to preserve the fusion rules of the anyons.

The importance of whether  the $\Z_k$ parafermion FQH states could indeed be realized in some experimental setup stems from 
the fact that their quasiparticle excitations are truly non-Abelian and therefore the exchanges  of their coordinates could 
generate non-diagonal braid matrices multiplying the quantum state's multiplet. These braid matrices can eventually be used as 
quantum gates for topological quantum computation \cite{kitaev-TQC,preskill-TQC,sarma-RMP}. 
Therefore, the interferometric patterns of the $\Z_k$ parafermion 
FQH states, expressed in terms of the analytic $S$ matrix elements, could provide the necessary signature for experimental identification
of non-Abelian anyons, opening in this way new perspectives for the implementation of universal topological quantum computers
 \cite{freedman-larsen-wang-TQC}.
 
\begin{figure}[htb]
\centering
\includegraphics[viewport=40 220 560 620,width=7cm,clip]{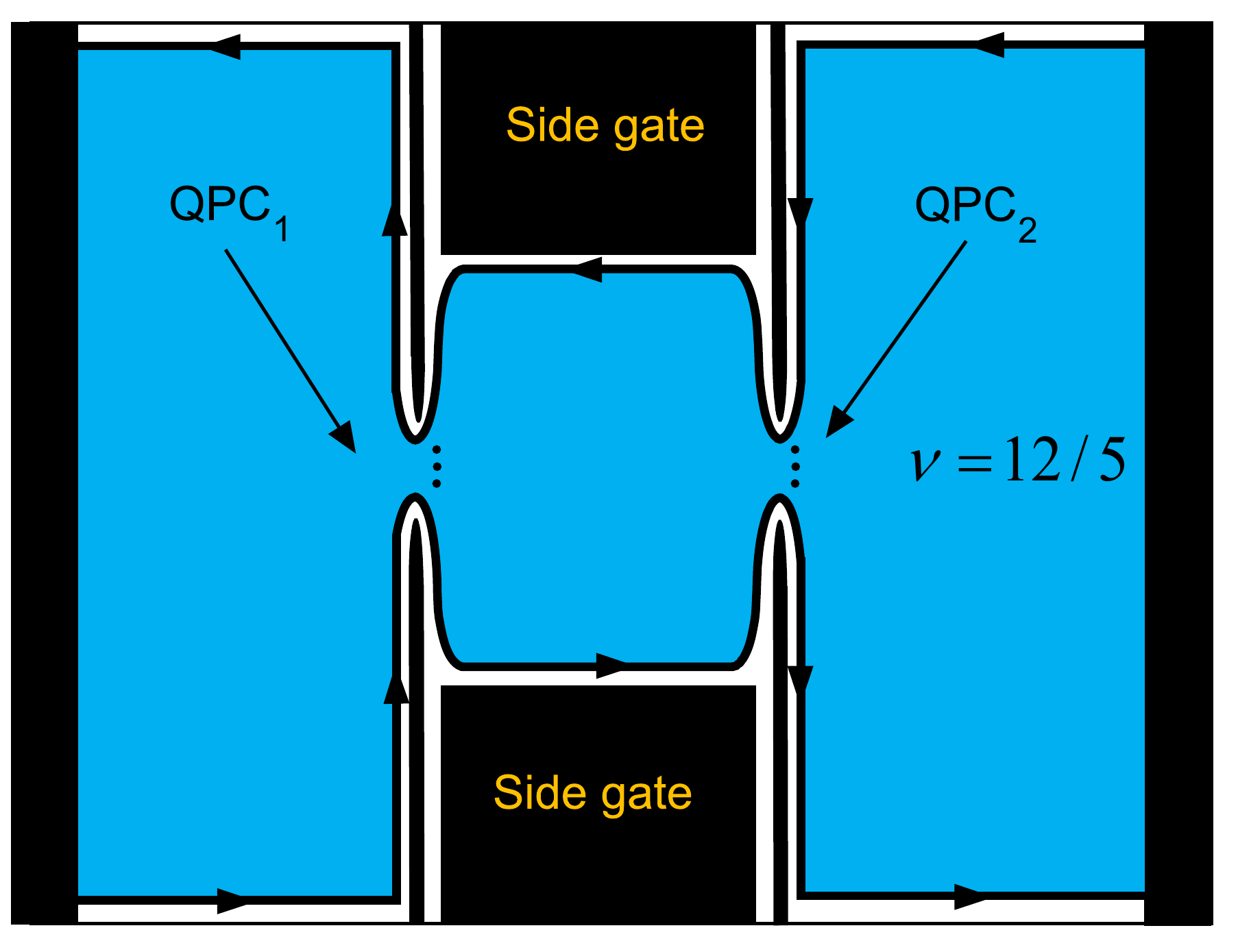}
\caption{Fabry--P\'erot interferometer. Two quantum point contacts inside of a FQH bar, with a filling factor $\nu_H=12/5$ as an example,
 create a precursor of a Coulomb-blockaded island whose area, respectively magnetic flux, could be varied by changing 
the voltage of the side gates.}
\label{fig:FPI}
\end{figure}

For the analysis of the interference patterns in Fabry--P\'erot interferometers we will be interested in the backscattered current  
in the weak-backscattering  regime shown in Fig.~\ref{fig:FPI}
where the two quantum point contacts (QPC), denoted as   QPC$_1$ and QPC$_2$,  are not completely pinched off and there is only a 
small current 
reflected back from the interferometer, due to the tunneling of quasiparticles through the QPCs, while almost all electric charge is 
transmitted through the interferometer along the edge states denoted by arrows in Fig.~\ref{fig:FPI}.
To lowest order in the amplitudes $t_1$ and $t_2$, for tunneling of fundamental quasiparticles through quantum-point contacts 
QPC$_1$ and QPC$_2$ respectively, the  amplitude of the backscattered current of the Fabry--P\'erot  interferometer shown
 in Fig.~\ref{fig:FPI}, in the quantum state $|\Psi\ra$ of the strongly correlated  FQH electron system, is proportional 
to the ``diagonal'' conductivity \cite{bonderson-5-2}
\beqa \label{s_xx}
\s_{xx}  & \propto & || \left( t_1 U_1 + t_2 U_2 \right) |\Psi\ra ||^2 = 
\la \Psi |   (t_1^* U_1^\dagger + t_2^* U_2^\dagger)\left( t_1 U_1 + t_2 U_2\right) |\Psi\ra \nn
& = & |t_1|^2 + |t_2|^2
 +2 \mathrm{Re} \left( t_1^* t_2 \la \Psi |   U_1^{-1} U_2 |\Psi\ra \right). 
\eeqa
The matrix element appearing in Eq.~(\ref{s_xx}) of the two unitary operators $U_1$ and $U_2$, each of which represents  
the quasiparticle evolution in the  state $|\Psi\ra$  during the process of tunneling  
through QPC$_1$ and  QPC$_2$ respectively,  determines the interference effects and can be written as \cite{bonderson-5-2}
\beq\label{alpha}
 \la \Psi |   U_1^{-1} U_2 |\Psi\ra = \ex^{i\alpha}  \la \Psi |  ( B_1)^{2}|\Psi\ra  = \ex^{i\alpha}  \la \Psi |   M |\Psi\ra ,
\eeq
where $\alpha$ is an Abelian phase which is a sum of the dynamical phase associated with the unitary evolution of the quasiparticle
 transported
along the full path around the central region (the island) of the interferometer containing $n$ fundamental quasiparticles and  the 
topological phase due to the Aharonov--Bohm effect of the electrically charged quasiparticles in the total magnetic field. The expectation 
value of $(B_1)^2\equiv M$  represents only the action of the pure braiding operator taking the traveling quasiparticle around the 
static quasiparticles localized  in the central region. While for Abelian quasiparticle of type $a$ transported along a complete loop around 
quasiparticle of type $b$ this monodromy always
satisfies  $| \la \Psi |   M_{ab} |\Psi\ra|=1$, the monodromy expectation value  for non-Abelian anyons  $| \la \Psi |   M_{ab} |\Psi\ra|\leq 1$ 
and could eventually be 0, which corresponds to no interference at all \cite{stern-halperin-5-2,bonderson-5-2}. 
This might  provide a promising experimental signature for the detection of non-Abelian quasiparticles if they
appear to be realized in some experimental setup.

Remarkably enough, for the lowest order tunneling interference process, the monodromy expectation value for a 
quantum state $|\Psi_{ab}\ra$ of uncorrelated quasiparticles of type $a$ and $b$, can be computed exactly \cite{bonderson-12-5}  in 
terms of the  modular $S$ matrix \cite{CFT-book} according to
\beq \label{M}
 \la \Psi_{ab} |   M |\Psi_{ab}\ra = \frac{S_{ab} S_{00}}{S_{0 a} S_{0b}},
\eeq
where $S_{ab}$ is the matrix element of the modular $S$ matrix corresponding to the topological charges $a$ of the quasiparticle being 
transported along a complete loop around a quasiparticle of topological charge $b$, while $0$ labels the vacuum sector, 
i.e., the state without any quasiparticle. Therefore if we know the $S$ matrix explicitly we can compare all interference patterns
 corresponding
to given types of static quasiparticles localized in the central region of the interferometer trying in this way to extract information about 
monodromy matrix elements and proving or disproving the emergence of non-Abelian quasiparticles in each experimental setup.

The main result in this paper is the explicit derivation of a compact analytic formula  for the full $S$ matrix for 
the $\Z_k$ parafermion quantum Hall droplets
using the level--rank duality \cite{CFT-book} between the affine Lie algebras $\widehat{su(k)_2}$ and $\widehat{su(2)_k}$.
The neutral part of the CFT, which represents  the $\Z_k$ parafermions, has been realized in Ref.~\cite{NPB2001} as a diagonal affine coset 
construction from a family of special maximally symmetric Abelian Hall states, denoted as 
\beq\label{symbol}
(L| {}^{\o}\Gamma_{W})=(3 \; | \; {}^{\L_1}A_{k-1} \ {}^{\L_1}A_{k-1})
\eeq
 in the notation of Ref.~\cite{fro-stu-thi}, which are rational conformal field theory extensions of the $\uu^{2k-1}$ current algebra \cite{kt}  
with vertex exponents \cite{kt} whose 
charges form an $(2k-1)$-dimensional integer lattice $\G$ \cite{fro-stu-thi}. The Gram matrix $G_\G$, or the metrics of the charge lattice, 
can be written in an appropriate basis as \cite{NPB2001}
 \beq\label{G}
G_\G =\left[
\begin{array}{c|c|c}
3 &  1 \; 0  \cdots  \ 0 & 1 \;  0  \cdots \ 0\cr \hline
\begin{array}{c}
1 \cr 0 \cr \vdots \cr 0
\end{array} &  C_{k-1} & 0 \cr \hline
\begin{array}{c}
1\cr 0 \cr \vdots \cr 0
\end{array} &  0 & C_{k-1}
\end{array} \right],
\eeq
where $C_{k-1}$ is the Cartan matrix of the Lie algebra $su(k)\equiv A_{k-1}$ and  the corresponding filling factor
can be computed from the so-called charge vector in the dual lattice $Q^T=(1,0, \ldots , 0) \in \G^*$ by  \cite{NPB2001}
\[
 \nu_H =Q^T. G_\G^{-1}.Q= \frac{k}{k+2} ,
\]
which coincides with the filling factor (\ref{nu}). This seems to be the minimal possible maximally symmetric chiral quantum Hall lattice 
for the filling factor (\ref{nu}) within the classification scheme of  Ref.~\cite{fro-stu-thi}. The first basis vector, corresponding to the diagonal 
element $(G_\G)^{11}=3$ in Eq.~(\ref{G}), represents the charged $\uu$ sector responcible for the Aharonov--Bohm effect. 
The rest of the basis vectors represent the neutral sector of this rational CFT model. This Abelian parent of the $\Z_k$ parafermion
FQH states possesses an $SU(k)\times SU(k)$ symmetry in its neutral sector as can be seen from the block-diagonal structure of the 
Gram matrix marked by the horizontal and vertical lines in Eq.~(\ref{G}). 
Its chiral algebra contains as a subalgebra the $\widehat{su(k)_1}\oplus \widehat{su(k)_1}$ Kac--Moody algebra and the 
$\Z_k$-parafermion CFT has been constructed as a diagonal affine coset \cite{gko,CFT-book,NPB2001} by projecting out its diagonal 
subalgebra $\widehat{su(k)_2}$ generated by the sums of the currents in both copies of  $\widehat{su(k)_1}$ \cite{NPB2001} 
\beq\label{PF_k}
\PF_k=
\frac{\widehat{su(k)_1}\oplus \widehat{su(k)_1}}{\widehat{su(k)_2}}.
\eeq
The CFT of the diagonal coset  (\ref{PF_k}) corresponds to Virasoro central charge \cite{CFT-book}
\beq\label{c_PF}
c_{\PF_k}=c_{\widehat{su(k)_1}} + c_{\widehat{su(k)_1}} -
c_{\widehat{su(k)_2}} =\frac{2(k-1)}{(k+2)} .
\eeq
The derivation of the neutral sector $S$ matrix contains two steps: first we construct the diagonal coset $S$ matrix 
from those for the numerator and the denominator of the coset  $\PF_k$ and then we use the 
level--rank duality \cite{CFT-book} between the affine Lie algebras $\widehat{su(k)_2}$ and $\widehat{su(2)_k}$ to
write the coset matrix in a more compact form. Remarkably, though seemingly accidental, the obtained explicit formulas 
prove that the modular $S$ matrix
of the diagonal coset precisely coincides with that for denominator $\widehat{su(k)_2}$ of the coset. 
Finally we use the full CFT $S$ matrix constructed in Ref.~\cite{NPB-PF_k} 
to calculate analytically the $S$-matrix elements and find the monodromy contribution of non-Abelian anyons to the interferometric 
patterns in Fabry--P\'erot interferometers.

The rest of this paper is organized as follows:   in Sect.~\ref{sec:ch} we describe the character decomposition of the chiral $\Z_k$
parafermion CFT in terms of charged- and neutral- partition functions and express the full $S$ matrix in terms of those for the 
charged and neutral sectors. In Sect.~\ref{sec:duality} we use the level--rank duality $\widehat{su(k)}_2 \simeq \widehat{su(2)}_k$ 
between the two affine Lie algebras to find a compact analytical form of the $S$ matrix for the denominator of the coset (\ref{PF_k}).
 In Sect.~\ref{sec:fibonacci} we give a detailed derivation of the $S$ matrix for the case $k=3$ correspondning to the Fibonacci 
anyons.   In Sect.~\ref{sec:fusion} we derive the fusion rules in the diagonal coset $\PF_k$ using its relation to the $su(2)_k/u(1)$
whose fusion rules are well known. Finally, in Sect.~\ref{sec:full} we give the compact analytic formula for the full $S$ matrix for 
the $\Z_k$ parafermion  (Read--Rezayi) FQH states.
\section{Full chiral parafermionic characters and  full modular $S$-matrix}
\label{sec:ch}
The full parafermionic CFT, including the charged  $\uu$ part of the Read--Rezayi states \cite{rr} and the neutral parafermionic part, 
which can be obtained by applying the diagonal coset projection (\ref{PF_k}) to the neutral part of the decomposable 
Abelian CFT \cite{NPB2001}, can be written symbolically as
\beq \label{full-CFT}
\left( \uu_{k(k+2)} \otimes \frac{\widehat{su(k)_1}\oplus \widehat{su(k)_1}}{\widehat{su(k)_2}} \right)^{\Z_k},
\eeq
where $\uu_{k(k+2)}$ represents the charged sector of the CFT and the subscript $k(k+2)$ means that the $\uu$ current algebra 
is rationally extended by a pair of vertex exponents  \cite{kt}  $\np{\ex^{\pm i \sqrt{k(k+2)} \phi(z)}}$ and the neutral component in the tensor product 
represents the $\Z_k$ parafermions realized as diagonal cosets. The superscript $\Z_k$,  over the tensor product of the $\uu$ and 
neutral  subalgebra described by the coset, represents the $\Z_k$ pairing rule given in Eq.~(3.16) in Ref.~\cite{NPB2001} 
between the charged representation spaces  labeled by the charge $l$ and the neutral spaces labeled by the coset weights
$\L_\mu+\L_\rho$  and can be expressed as follows: the only allowed excitations of the full $\Z_k$ parafermions CFT are those which 
satisfy the $\Z_k$ pairing rule
\beq \label{PR}
\mu+\rho = l \mod k, \quad \mathrm{where} \ \ 0 \le \mu \le \rho \leq k-1 \ \  \mathrm{and} \ \ l \mod k+2 .
\eeq
One of the advantages of the diagonal coset construction of the $\Z_k$ parafermions is that the $\Z_k$ pairing rule (\ref{PR})
of the full CFT is naturally inherited from a decomposition relation (see Eq.~(3.12) in Ref.~\cite{NPB2001})  between the original 
indecomposable Abelian parent CFT, described by the charge lattice denoted by (\ref{symbol})  in the notation of Ref.~\cite{fro-stu-thi}, 
and a decomposable sublattice Abelian CFT \cite{NPB2001} in which the neutral and charged 
degrees of freedom are completely decoupled, i.e., all combinations of $l$ and $\L_\mu+\L_\rho$ are allowed in the decomposable 
sublattice Abelian CFT,  but not in the original one, in which they have to satisfy~\cite{NPB2001} the restrictive paring rule (\ref{PR}).

After the diagonal coset projection \cite{NPB2001}, which can be denoted symbolically as
\[
 \widehat{su(k)_1}\oplus \widehat{su(k)_1} \to  \frac{\widehat{su(k)_1}\oplus \widehat{su(k)_1}}{\widehat{su(k)_2}},
\]
the total disk partition function (Grand canonical)
$\chi_{l,\rho} (\beta,\mu)= \tr_{\H_{l,\rho}}\exp\left( -\beta (H-\mu_{\ch} N) \right)$
for the $\Z_k$ parafermion quantum Hall islands \cite{NPB2001} can 
be labeled by two integers, $l \mod k+2$ and  $\rho\mod k$ satisfying $l-\rho \leq \rho \mod k$, corresponding to the Hilbert space 
$\H_{l,\rho}$, and can be written as follows
 \beq \label{full-ch}
\chi_{l,\rho} (\t,\z) = \sum_{s=0}^{k-1} K_{l+s(k+2)}(\t,k\z;k(k+2)) \ch(\L_{l-\rho+s}+\L_{\rho+s})(\t),
\eeq
where the modular parameters $\t$ and  $\z$ of the rational CFT are related to the inverse  temperature  $\beta$   \cite{CFT-book},
and chemical potential $\mu_{\ch}$, respectively, \cite{NPB2001,NPB2015} by
\beq\label{modular}
q=\e^{-\beta\D\varepsilon}=\e^{2\pi i \t}, \quad \D \varepsilon= \hbar\frac{2\pi v_F}{L}, \quad \z = \frac{\mu_{\ch}}{\D \varepsilon} \t ,
\eeq
with $v_F$ being the Fermi velocity on the disk's edge and $L$ the disc circumference. 
The explicit formulas for the $\uu$ contribution $K(\t,\z;m)$ and the neutral partition functions  $\ch(\L_\mu+\L_\nu;\t)$ 
can be found in Ref.~\cite{NPB2001}, however they will not be needed here and we skip them.

The modular $S$-matrix $S_{ij}$ for any rational CFT is defined in general as the transformation matrix for the characters $\chi_i$ 
 under the modular inversion \cite{CFT-book} $\t  \to -1/\t$, i.e.,
\[
\chi_{i} (-1/\t,\z/\t) = \sum_j S_{ij} \chi_{j} (\t,\z) .
\]
Recall that the two transformations
\[
T: \t \to \t+1, \quad S: \t  \to -1/\t, \quad (ST)^3=S^2=C,
\]
where $C^2=1$ is the charge conjugation, 
generate the modular group \cite{CFT-book} characterizing any rational CFT.

The modular $S$-matrix $\Sc^{\L}{}_{\L'} $ for the diagonal coset CFT has been derived in Appendix B in Ref.~\cite{NPB-PF_k} 
(see Eq. (59) there) in terms of 
the modular $S$-matrices of the numerator and denominator  of the coset (\ref{PF_k}), the character decomposition and 
the properties of the modular $S$ matrices under the action of the simple currents,
and can be written as
\beq\label{S-coset}
\Sc^{\L_{\mu}+\L_{\nu}}{}_{\L_{\rho}+\L_{\sigma}} = \ex^{2\pi i \frac{(\mu+\nu)(\rho+\sigma)}{k}}
\left(S^{(2)}_{\L_{\mu}+\L_{\nu}, \L_{\rho}+\L_{\sigma}}\right)^{*} , 
\eeq
where $\L_{\mu}+\L_{\nu}$ and $\L_{\rho}+\L_{\sigma}$ with $0\leq \mu  \leq \nu \leq k-1$ and $0 \leq \rho \leq \sigma \leq k-1 $
label two irreducible representation of the diagonal coset (\ref{PF_k}),
 $S^{(2)}_{\L_{\mu}+\L_{\nu}, \L_{\rho}+\L_{\sigma}}$ is the  modular $S$ matrix for the current algebra  $\widehat{su(k)_2}$
 and the $*$ means complex conjugation.

The  matrix    $S^{(2)}_{\L_{\mu}+\L_{\nu}, \L_{\rho}+\L_{\sigma}}$  could be computed by the Weyl--Kac formula \cite{kac}
\beq\label{S.2}
S^{(2)}_{\L,\L'} = \frac{i^{|\D^+|}}{\sqrt{|M^*/hM|}}
\sum_{w\in \mathcal{W}} \epsilon(w)
\exp\left( -\frac{2\pi i}{h} \left( \L+\underline{\rho}\vert w\left(\L'+\underline{\rho}\right)\right)\right)
\eeq
where $\D^+$ is the set of all positive roots (in this case for $su(k)$),
$M=\Z \, \alpha^1+\cdots + \Z \, \alpha^{k-1}$ is the root lattice,
$M^*=\Z \, \L^1+\cdots + \Z \, \L^{k-1}$ is its dual, $h=k+2$,
$\underline{\rho}$ is half the sum of the positive roots and
$\mathcal{W}$ is the Weyl group  ($\epsilon(w)$ is the determinant of the
Weyl reflection $w$ as a matrix in the basis of simple roots).
In this particular case
we have $h M \subset M \subset M^* \subset (h M)^*$ so that
\[
\left| M^* /h M \right|=\frac{\sqrt{\det (h M)}}{\det M^*} =
\frac{\sqrt{h^2 \det C}}{\sqrt{\det C^{-1}}} = h \det C = k(k+2)
\]
where $C$ is again the Cartan matrix for $su(k)$. \\

\section{Level--rank duality and (neutral) diagonal-coset $S$ matrix}
\label{sec:duality}
The  level--rank duality \cite{CFT-book} is in general an equivalence relation between different 
two-dimensional CFT models based on affine Lie algebras.
In particular we will be interested in the duality $\widehat{su(N)}_l \simeq \widehat{su(l)}_N$ 
where $N=k$ and $l=2$ in the notation of Ref.~\cite{CFT-book}, i.e. in our case of the diagonal coset we will consider the 
level--rank duality
\beq \label{duality}
\widehat{su(k)}_2 \simeq \widehat{su(2)}_k .
\eeq
One of the motivations to investigate this correspondence is that
as we have seen in Eq.~(\ref{S.2}), the modular $S$ matrix for the $\widehat{su(k)}_2 $ current algebra can be expressed as a 
sum over the Weyl group which contains $k!$ elements while the $S$ matrix for 
$\widehat{su(2)}_k $ contains only two elements and it can be conveniently written as \cite{CFT-book}
\beq\label{S-su2_k}
S^{\widehat{su(2)}_k}_{l, l'} =\sqrt{\frac{2}{k+2}} \sin\left(  \pi \frac{(l+1)(l'+1)}{k+2}\right) , \quad 0\leq l, l'\leq k.
\eeq
However, the correspondence between the integrable representations of $\widehat{su(N)}_l$ and $\widehat{su(l)}_N$ 
is not one-to-one \cite{CFT-book} as can be seen from their different numbers $\frac{(N+l-1)!}{l! (N-1)!}$ and $\frac{(N+l-1)!}{N! (l-1)!}$, 
respectively.
Nevertheless, if we divide both numbers by $N$ and $l$, respectively, these numbers become equal 
(cf. Eq.~(16.159) on p. 703 in \cite{CFT-book}). This division represents a factorization over the action of the corresponding centers $\Z_N$
for $\widehat{su(N)}_l$ and $\Z_l$ for $\widehat{su(l)}_N$ and the equivalence relation is $\l \simeq J^s(\l)$, where $J$ is the 
simple current generating the outer-automorphism group \cite{CFT-book}. Thus we conclude that in the level--rank duality 
there is a one-to-one correspondence between the orbits of the outer-automorphism groups of $\widehat{su(k)}_2$ and 
$\widehat{su(2)}_k $, which are represented by the action of the corresponding simple currents.

In addition to the one-to-one correspondence between the simple-current orbits of integrable representations
there is a remarkable relation between the modular $S$ matrices of the two affine Lie algebras related by the level--rank duality.
In more detail, if $\l$ and $\l'$ are the weights of two integrable representations of $\widehat{su(k)}_2$ then the 
$S$ matrix of $\widehat{su(k)}_2$ 
can be expressed in terms of that for $\widehat{su(2)}_k $ as follows \cite{CFT-book}
\beq \label{S-duality}
S^{\widehat{su(k)}_2}_{\l, \l'} =\sqrt{\frac{2}{k}}\ \ex^{2\pi i \frac{|\l|.|\l'|}{2k}} \ S^{\widehat{su(2)}_k}_{{}^t\!\l, {}^t\!{\l'}} ,
\eeq
where $|\l|$ and $|\l'|$ are the number of boxes in the reduced Young tableaux  corresponding to  the integrable representations 
with weights $\l$ and $\l'$, respectively, while ${}^t\!\l$, ${}^t\!{\l'} $ denote the $\widehat{su(2)}_k $ integrable representations whose 
weights correspond to the transposed Young tableaux corresponding to $\l$ and $\l'$,  respectively, 
cf. Eq.~(16.168) on p. 705 in \cite{CFT-book}.

In order to obtain the $S$-matrix elements within each orbit of the simple currents we can use the general property of the $S$ matrices
under the action of simple currents \cite{CFT-book,schw,schw-CMP} 
\beq \label{S_J}
S_{J*(\L),\L'}=\ex^{-2\pi i \tilde{Q}_J(\L')} S_{\L,\L'},
\eeq
where $J*(\L)$ is the  weight  obtained after the action of the simple current $J$ over the weight $\L$ and
the monodromy 
charge\footnote{The monodromy charge $\tilde{Q}_J(\L)$ of the weight $\L$ with respect to the simple current 
$J$, defined in Eq.~(\ref{mon-3}), differs by a minus sign from that defined in Refs. \cite{schw,schw-CMP} and the 
sign in the exponent in Eq.~(\ref{S_J}) is respectively the opposite.}
\beq \label{mon-3}
\tilde{Q}_J(\L) =\Delta (J*(\L)) - \Delta (\L) - \Delta (J)  \mod \Z.
\eeq
is expressed in terms of the CFT dimensions $\Delta$ of the chiral conformal fields labeled by the weight $\L$,  $J*(\L)$ 
and the CFT dimension of the simple current $J$.

We shall continue with the explicit computation of the modular $S$ matrix for $\widehat{su(k)}_2$  in terms the elements of the $S$ matrix 
(\ref{S-su2_k}). The list of integrable weights for $\widehat{su(k)}_2$,  labeled by $\L_\mu+\L_\nu$ with $0\leq \mu \leq \nu \leq k-1$, splits
into orbits of simple current's action. The simple current $J$ for $\widehat{su(k)}_2$ acts on the corresponding weights by fusion as
\[
J*(\L_\mu+\L_\nu)= (\L_1+\L_1)\times (\L_\mu+\L_\nu)=\L_{\mu+1}+\L_{\nu+1}.
\]
Therefore, each weight $\L_\mu+\L_\nu$ can be obtained from some fundamental weight $\L_0+\L_l$  by the repeated 
action of the simple current $J$ as
\[
(\L_\mu+\L_\nu) = J^{\mu}*(\L_0+\L_{\nu-\mu}) \quad \mathrm{with} \quad 0\leq \nu-\mu \leq k-1 .
\]
Thus, we conclude that the orbits under the action of the simple currents can be represented 
by $\L_0+\L_l$ with $l=\nu-\mu$ as 
\beqa\label{orbits-suk_2}
\left[ \ \L_0+ \L_0 \  \right]  &=&  \left\{ \ \L_0+ \L_0,  \ \L_1+ \L_1, \ldots , \L_{k-1}+ \L_{k-1} \ \right\}  \cr
\left[ \ \L_0+ \L_1 \  \right]   &=& \left\{ \ \L_0+ \L_1, \  \L_1+ \L_2, \ldots , \L_{k-2}+ \L_{k-1}  \ \right\} \cr
& \vdots & \cr
\left[ \ \L_0+ \L_{r} \  \right]   &=& \left\{ \ \L_0+ \L_{r}, \  \L_1+ \L_{r+1}, \ldots  ,\L_{r}+ \L_{2r}  \ \right\} ,
\eeqa
where it is understood that $\L_\beta +\L_\alpha = \L_\alpha +\L_\beta$ if $\beta \geq \alpha$ and $\L_{k+s} =\L_s$.
In other words, the orbits can be represented by (some of) the fundamental weights $\L_0+\L_l$  and 
$\L_\mu+\L_\nu \in [\L_0+\L_{\nu-\mu}]$.
The number $r$ of orbits can be determined from the condition $2r \leq k$ because the elements in the orbit of $\L_0+ \L_{r}$ 
obtained  by the action of the simple current $J$ start repeating the elements of the previous orbits, i.e.,
$J*(\L_r+ \L_{2r}) \in [\L_0+ \L_{i}]$ with $0\leq i \leq r$ when $2r=k (= 0 \mod k)$ if $k$ is even or $2r=k+1$ if $k$ is odd, i.e.
\beq \label{orbit-r}
r=\left\{ \begin{array}{cl} k/2+1 & \mathrm{if} \quad k=\mathrm{even} \cr (k+1)/2 & \mathrm{if} \quad k=\mathrm{odd}  \end{array} .\right.
\eeq

On the other hand, we have to consider the  orbits of  the simple current's action over the integrable 
representations of $\widehat{su(2)}_k$, which are labeled by $l$, corresponding to the weights $\L_l$  
with CFT dimensions $\Delta(\L_l)$ that can be written  in the form
\beq\label{L_l}
\L_l=l\L_1, \quad \Delta(\L_l)=\frac{l(l+2)}{4(k+2)}, \quad \mathrm{where} \quad 0\leq l \leq k ,
\eeq
where $\L_1$ is the fundamental weight of $su(2)$.
The simple current of $\widehat{su(2)}_k$  corresponds to the $l=k$ primary field $J=\phi_k$ and has CFT dimension
$\Delta(\L_k)=k/4$ for general $k$. Its quantum dimension can be obtained form Eq.~(\ref{S-su2_k}) as follows
\[
{\mathcal D}_k=\frac{S^{\widehat{su(2)}_k}_{0,k}}{S^{\widehat{su(2)}_k}_{0,0}}=
\frac{\sin\left(\frac{(k+1)\pi}{k+2} \right)}{\sin\left(\frac{\pi}{k+2}\right)}=1.
\]
Therefore, the fusion rules of the $\widehat{su(2)}_k$ primary fields $\phi_l$ with the simple current $J$ 
could contain only one term in the right-hand side \cite{CFT-book}.
Thus, the general $\widehat{su(2)}_k$ fusion rules, which can be obtained from the Verlinde formula \cite{CFT-book},
\beq \label{fusion-su2_k}
\phi_l * \phi_{l'} = \mathop{\bigoplus}\limits_{l'' = |l-l'|}^{\min(l+l', 2k-l-l')}\phi_{l''}
\eeq
imply that the  fusion rules of the primary fields $\phi_l$ with the simple current $J=\phi_k$ are
\beq \label{J-su2_k}
J* \phi_l = \phi_{k-l}, \quad 0 \leq l \leq k,
\eeq
hence, $J^2=1$. 
Because of that the orbits of the integrable irreducible representations of $\widehat{su(2)}_k$ under the action of the 
simple current $J$ could contain at most two weights and can be written as
\beqa
 [ 0 ]  &=&  \left\{  0, k \right\}  \cr
 [ 1 ]  &=&  \left\{ 1, k-1 \right\} \cr
 &\vdots& \cr
 [ r ]  &=&  \left\{ r, k-r \right\} . \label{orbits-2}
\eeqa
Obviously, the number of orbits $r$ can be  determined from the condition $r \leq k-r$, or, equivalently from
$2r \leq k$ which gives  the same $r$ as in Eq.~(\ref{orbit-r}).

Now,  we can formulate more precisely the level--rank duality (\ref{duality}) as the one-to-one correspondence between the orbits 
in Eq.~(\ref{orbits-suk_2}) and  those in Eq.~(\ref{orbits-2}) according to
\beq \label{corr}
[\L_\mu+\L_\nu] \quad  \Longleftrightarrow \quad [l], \qquad \mathrm{with} \quad l = \nu -\mu \mod k.
\eeq

Next we will compute the $\widehat{su(k)}_2$ modular $S$ matrix for the orbits representatives  $\L_0+\L_{l}$. 
The integrable representation of $\widehat{su(k)}_2$ labeled by $\L_0+\L_l$ is the fundamental representation corresponding to the 
Young tableau consisting of $l$ boxes in one column. Therefore, the number of boxes $|\l|$ entering Eq.~(\ref{S-duality}) for
$\l=\L_0+\L_{l}$ is   $\left| \L_0+\L_{l} \right| = l$. At the same time, the transposed \index{Young tableau} Young tableau specifying the 
representation of $\widehat{su(2)}_k$  consists of one row containing $l$ boxes, which is exactly the integrable representation
given in Eq.~(\ref{L_l}). Therefore, the $\widehat{su(k)}_2$ modular $S$ matrix elements for  the orbits representatives are
 \beqa \label{S-orbits}
 S^{\widehat{su(k)}_2}_{\L_0+\L_{\nu-\mu}, \L_0+\L_{\sigma-\rho}} =
\sqrt{\frac{2}{k}}\ \ex^{2\pi i \frac{ \left| \L_0+\L_{\nu-\mu} \right| . \left|\L_0+\L_{\sigma-\rho} \right |}{2k}} \ 
S^{\widehat{su(2)}_k}_{\nu-\mu,  \sigma -\rho}  \nn
= \frac{2}{ \sqrt{k(k+2)} } \ 
\ex^{2\pi i \frac{ (\nu-\mu) (\sigma-\rho)}{2k}} 
\sin\left( \frac{\pi(\nu-\mu+1)(\sigma-\rho+1)}{k+2}\right). \quad \quad 
 \eeqa
In order to obtain the $S$ matrix elements for the other weights within each orbit we use the property (\ref{S_J}) of the $S$ 
matrix under the action of 
the simple currents. The monodromy charge $\tilde{Q}_J^\mu (\L_\rho+\L_\sigma)$ for the action of a simple current on the 
integrable weights  for $\widehat{su(k)}_2$ can be computed from the definition (\ref{mon-3}) 
\beq\label{mon-2.2}
\tilde{Q}_{J^{\mu}} (\Lambda_{\rho}+\Lambda_{\sigma}) =-\frac{\mu(\rho+\sigma)}{k} 
\eeq
where we used the following formula for the CFT dimension of the $\widehat{su(k)}_2$ representation labeled by $\L_{\mu}+\L_{\nu}$
\beq \label{D-2}
\Delta^{(2)} (\L_{\mu}+\L_{\nu}) = \frac{2\mu(k-\nu)+(k+1)(\mu(k-\mu)+\nu(k-\nu))}{2k(k+2)} .
\eeq
Thus we have 
\[
\left[\ex^{-2\pi i \tilde{Q}_{J^{\mu}} (\Lambda_{\rho}+\Lambda_{\sigma})} \right]^{-1} = \ex^{-2\pi i \frac{\mu(\rho+\sigma)}{k}} .
\]
so that we can write 
 \beqa
 S^{\widehat{su(k)}_2}_{\L_\mu+\L_\nu, \L_0+\L_{\sigma-\rho}}  &=&
S^{\widehat{su(k)}_2}_{J^\mu*(\L_0+\L_{\nu-\mu}), \L_0+\L_{\sigma-\rho}}  \nn 
&=&  \ex^{-2\pi i \tilde{Q}_{J^\mu} (\L_0+\L_{\sigma-\rho})} 
 S^{\widehat{su(k)}_2}_{\L_0+\L_{\nu-\mu}, \L_0+\L_{\sigma-\rho}} \nonumber
 \eeqa
 and similarly
 \beqa
 S^{\widehat{su(k)}_2}_{\L_\mu+\L_\nu, \L_\rho+\L_\sigma} &=& 
 S^{\widehat{su(k)}_2}_{\L_\mu+\L_\nu, J^\rho*(\L_0+\L_{\sigma-\rho})} \nn
&=&
 \ex^{-2\pi i \tilde{Q}_{J^\rho} (\L_\mu+\L_\nu )} 
 S^{\widehat{su(k)}_2}_{\L_\mu+\L_{\nu}, \L_0+\L_{\sigma-\rho}} . \nonumber
 \eeqa
Next, using Eq.~(\ref{mon-2.2})  we have
\[
\tilde{Q}_{J^\mu} (\L_0+\L_{\sigma-\rho}) = -\frac{\mu(\sigma-\rho)}{k}, \quad
\tilde{Q}_{J^\rho} (\L_\mu+\L_\nu )= -\frac{\rho(\mu+\nu)}{k}, \
\]
and using Eq.~(\ref{S-orbits}) we find 
 \beqa
 S^{\widehat{su(k)}_2}_{\L_\mu+\L_{\nu}, \L_\rho+\L_{\sigma}} & = &
 \frac{2}{ \sqrt{k(k+2)} } \ 
\ex^{2\pi i \left[ 
 \frac{ (\nu-\mu) (\sigma-\rho)}{2k} 
+\frac{\mu(\sigma-\rho)}{k} 
+\frac{\rho(\mu+\nu)}{k} \right]
} \nn  &\times &
\sin\left( \frac{\pi(\nu-\mu+1)(\sigma-\rho+1)}{k+2}\right) .
 \eeqa
After taking into account that 
 \[
  \frac{ (\nu-\mu) (\sigma-\rho)}{2k} +\frac{\mu(\sigma-\rho)}{k} +\frac{\rho(\mu+\nu)}{k} =
   \frac{ (\mu+\nu) (\rho+\sigma)}{2k}
 \]
we finally arrive at the following expression for the modular $S$ matrix of the $\widehat{su(k)}_2$ current algebra
\beqa \label{S-suk_2}
 S^{\widehat{su(k)}_2}_{\L_\mu+\L_{\nu}, \L_\rho+\L_{\sigma}} &=&
 \frac{2}{ \sqrt{k(k+2)} } \ 
 \ex^{2\pi i  \frac{ (\mu+\nu) (\rho+\sigma)}{2k}} \times \nn
& &\sin\left( \frac{\pi(\nu-\mu+1)(\sigma-\rho+1)}{k+2}\right) ,
 \eeqa
where $0\leq \mu \leq \nu \leq k-1$ and $0\leq \rho \leq \sigma \leq k-1$. \\

\noindent
\textbf{Remark:} 
\textit{The biggest advantage of Eq.~(\ref{S-suk_2}) is that  it contains only one term in contrast to the Weyl formula 
for the modular $S$ matrix of $\widehat{su(k)}_2$  given in Eq.~(\ref{S.2}) which contains $k!$ terms.
}\\

Now that we have the compact form (\ref{S-suk_2}) for the denominator of the diagonal coset
we can use the expression (\ref{S-coset}) for the coset $S$ matrix in terms of the $S$ matrix for $\widehat{su(k)}_2$,
denoted there as $S^{(2)}_{\L_\mu+\L_{\nu}, \L_\rho+\L_{\sigma}}$, taking into account the complex conjugation of the latter and 
combining the phase factors, to 
 finally obtain the diagonal-coset's modular $S$ matrix in the form
\beqa\label{Sc-2}
\dot{S}^{\L_\mu+\L_\nu}{}_{\L_\rho +\L_\s} &=&\frac{2}{\sqrt{k(k+2)}} \ \ex^{2\pi i \frac{(\mu+\nu)(\rho+\s)}{2k}} \times \nn
 & &\sin\left(  \pi \frac{(\nu-\mu+1)(\s-\rho+1)}{k+2}\right) . 
\eeqa
Comparing the expressions in Eqs.~(\ref{Sc-2}) and (\ref{S-suk_2}) we conclude 
that \textbf{the diagonal-coset's modular $S$ matrix is completely identical to that of $\widehat{su(k)}_2$}, i.e.,
\beq \label{equiv-S}
\dot{S}^{\L_\mu+\L_\nu}{}_{\L_\rho +\L_\s} \equiv S^{\widehat{su(k)}_2}_{\L_\mu+\L_{\nu}, \L_\rho+\L_{\sigma}}, 
\quad 
\left\{ \begin{array}{c}   0\leq \mu \leq \nu \leq k-1 \\ 0\leq \rho \leq \sigma \leq k-1 .
 \end{array} \right.
\eeq
\noindent
\textbf{Remark:} 
\textit{Although the relation between the diagonal coset (\ref{PF_k}) and the $\widehat{su(2)}_k/ \widehat{u(1)}$ realization 
of the $\Z_k$ parafermions \cite{zf-para,gep-qiu} is well studied, so that such a relation between the corresponding $S$ matrices 
might have been suspected, to our knowledge this is the first rigorous, compact and analytic derivation, based on the level--rank duality, of the 
diagonal-coset's $S$ matrix in the basis of the diagonal-coset weights $\L_\mu+\L_\nu$}.\\

The compact form of the diagonal-coset's $S$ matrix given in  Eq.~(\ref{Sc-2}) could be used 
for a direct analytic computation of the monodromy matrix for Fabry--P\'erot interferometer experiments for 
Fibonacci anyons \cite{bonderson-12-5} as we will see in the next Section.

\section{Example: $S$ matrix for the $\Z_3$ parafermions and measurement of Fibonacci anyons}
\label{sec:fibonacci}
As an interesting illustration of the general procedure, as well as a case of special importance for topological quantum 
computation with \index{Fibonacci anyons} Fibonacci anyons \cite{fibonaci}, we will consider below in more detail the example of $k=3$. 

The Fibonacci anyons \cite{fibonaci} can be realized in the diagonal coset of the 
$\Z_3$ parafermion FQH states \cite{NPB2001,NPB2015-2} (or, in the three-state Pots model) as the the parafermion primary 
fields $\varepsilon$ corresponding to the nontrivial orbit of the simple-current's action, i.e., 
\beqa\label{fibonacci}
\I=\left[ \ \L_0+ \L_0 \  \right]  &=&  \left\{ \ \L_0+ \L_0,  \ \L_1+ \L_1, \  \L_2+ \L_2 \ \right\}  \cr
\varepsilon=\left[ \ \L_0+ \L_1 \  \right]   &=& \left\{ \ \L_0+ \L_1, \  \L_1+ \L_2, \ \L_0+ \L_2  \ \right\} .
\eeqa
 with the following fusion rules
\[
\I \times \I = \I , \quad \I \times \varepsilon = \varepsilon, \quad \varepsilon \times \varepsilon = \I +\varepsilon .
\]
The information encoding for Fibonacci anyons is in the fusion channels, denoted again by the fields $\I$ and $\varepsilon$ of the resulting 
fusion channel, i.e., $(\varepsilon,\varepsilon)_{\I}$ means that if the pair of Fibonacci anyons is fused the resulting field will be $\I$, while 
$(\varepsilon,\varepsilon)_{\varepsilon}$ means that if the pair of Fibonacci anyons is fused the resulting field will be $\varepsilon$. 
The fusion channel is a robust topological characteristics of the anyon pair which is preserved even when the two anyons in the pair are 
well separated. Now, for triples of anyons we have the following definition of the computational basis \cite{fibonaci}
\beqa\label{comp}
|0\ra &=&((\varepsilon,\varepsilon)_{\I} , \varepsilon)_{\varepsilon}  \nn
|1\ra &=&((\varepsilon,\varepsilon)_{\varepsilon}, \varepsilon)_{\varepsilon}   ,
\eeqa
which means that if the first pair in the triple fuses to $\I$ this is the state $|0\ra$ while if the first pair in the triple fuses to $\varepsilon$ this is 
the state $|1\ra$. The quantum dimension of the states in the computational basis (\ref{comp}) is $(1+\sqrt{5})/2$. 
There is one also a third state for three Fibonacci anyons $((\varepsilon,\varepsilon)_{\varepsilon}, \varepsilon)_{\I}$, 
having a trivial quantum dimension, which decouples from the previous two and is called non-computational  \cite{fibonaci}.

Next, ordering the $\widehat{su(3)}_2$  integrable weights, defining the basis for the $S$ matrices for  both
$\widehat{su(3)}_2$ and the diagonal coset $\PF_3$ defined in Eq.~(\ref{PF_k}) for $k=3$, as follows
\[
\left[ \ \L_0+\L_0, \   \L_1+\L_1, \ \L_2+\L_2, \  \L_0+\L_1, \  \L_0+\L_2, \ \L_1+\L_2   \  \right],
\]
we can write the modular matrix  $\dot{S} \equiv S^{\widehat{su(3)}_2}$ of the diagonal coset $\PF_3$ 
in the following form (see \ref{app} for more details)
\beq \label{S-su3_2}
\dot{S}= \frac{1}{\Dc} 
\left[\matrix{
1 & 1 & 1  & \delta & \delta & \delta \cr
1 & \ex^{-\frac{2\pi i}{3}} & \ex^{\frac{2\pi i}{3}} & \ex^{\frac{2\pi i}{3}}\delta & \ex^{-\frac{2\pi i}{3}}\delta & \delta \cr
1 & \ex^{\frac{2\pi i}{3}} & \ex^{-\frac{2\pi i}{3}} & \ex^{-\frac{2\pi i}{3}}\delta & \ex^{\frac{2\pi i}{3}}\delta & \delta \cr
\delta & \ex^{\frac{2\pi i}{3}}\delta & \ex^{-\frac{2\pi i}{3}}\delta & \ex^{\frac{i\pi }{3}} & \ex^{-\frac{i\pi }{3}} & -1 \cr
\delta & \ex^{-\frac{2\pi i}{3}}\delta & \ex^{\frac{2\pi i}{3}}\delta & \ex^{-\frac{i\pi }{3}} & \ex^{\frac{i\pi }{3}} & -1 \cr
\delta & \delta & \delta & -1 & -1 & -1 \cr
} \right],
\eeq
where $\delta$ is the Golden ratio defined in Eq.~(\ref{delta}) and $\Dc$ is the total quantum dimension defined in Eq.~(\ref{Dc}).
This matrix is the same (up to complex conjugation) as that obtained in Ref.~\cite{bonderson-12-5}, 
however, the authors use the equivalence of the diagonal coset $\PF_3$ to the coset $\widehat{su(2)}_3/\widehat{u(1)}$, which we 
will consider in more detail in the next Section, where we will establish precisely the one-to-one correspondence between the two cosets. 
Here we derived this $S$ matrix by using the level--rank duality which is the  more direct approach.
Nevertheless it is important to remember that the diagonal-coset realization of the $\Z_k$ parafermion FQH states is more natural,
provides a mechanism to inherit the $\Z_k$ pairing rule from the Abelian parent CFT and provides a way to construct the manybody 
wave functions by symmetrization of the correlation functions of the Abelian parent CFT \cite{NPB2001}, which are very convenient for
numerical calculations.

Now we will show how to use the explicit result  (\ref{S-su3_2}) obtained for the diagonal-coset $S$ matrix to identify the eventual presence of
Fibonacci anyons in a Fabry--P\'erot interferometer described in Fig.~\ref{fig:FPI} using the monodromy computed from Eq.~(\ref{M}) 
in terms of the $S$-matrix elements. Imagine that there is no Fibonacci anyon $\varepsilon$ in the island formed between the two
QPCs and let us assume that the lowest order tunneling process between the QPCs is realized by a Fibonacci anyon. 
Then the quantum state of the system can be denoted as $|\I, \varepsilon \ra$, i.e., $a=\L_{0}+\L_{0}$,  $b=\L_{0}+\L_{1}$
so that the monodromy matrix element entering Eq.~(\ref{M}) would be 
\[
\la \I,  \varepsilon | M | \I, \varepsilon \ra = 
\frac{\dot{S}^{\L_{0}+\L_{0}} {}_{\L_0+\L_1}  \  \dot{S}^{\L_{0}+\L_{0}} {}_{\L_0+\L_0}     }{\dot{S}^{\L_{0}+\L_{0}} {}_{\L_0+\L_0}  \ \dot{S}^{\L_{0}+\L_{0}} {}_{\L_0+\L_1} } =1 .
\]
If, on the other hand there is a Fibonacci anyon in the vicinity of the Fabry--P\'erot interferometer the quantum state of the system
would be $|\varepsilon, \varepsilon \ra$, i.e., $a=\L_{0}+\L_{1}$,  $b=\L_{0}+\L_{1}$
so that the monodromy matrix element entering Eq.~(\ref{M}) would be 
\[
\la \varepsilon, \varepsilon | M | \varepsilon, \varepsilon \ra = 
\frac{\dot{S}^{\L_{0}+\L_{1}} {}_{\L_0+\L_1}  \  \dot{S}^{\L_{0}+\L_{0}} {}_{\L_0+\L_0}     }{\dot{S}^{\L_{0}+\L_{0}} {}_{\L_0+\L_1}  \ \dot{S}^{\L_{0}+\L_{0}} {}_{\L_0+\L_1} } = \frac{\dot{S}_{46}  \dot{S}_{11}}{\dot{S}_{14}  \dot{S}_{16}}=-\frac{1}{\delta^2} ,
\]
where $\dot{S}_{ij}$ is the matrix element of $\dot{S}$ on the $i$-th row and $j$-th column in Eq.~(\ref{S-su3_2}).
Because the matrix element $\la \varepsilon, \varepsilon | M | \varepsilon, \varepsilon \ra$ multiplies $\e^{i\alpha}$ in Eq.~(\ref{alpha}),
where $\alpha$ contains the Aharonov--Bohm interference term when the magnetic field is varied, the interference in the second case, when
a Fibonacci anyon is present in the Fabry--P\'erot interferometer is suppressed by the factor $1/\delta^2 \approx 0.38$ compared to the case
when there is no Fibonacci anyon in the interferometer. In general, for Abelian anyons 
$|\la \Psi | M | \Psi \ra|=1$ so the suppression of the Aharonov--Bohm interference is a clear sign
of the presence of non-Abelian anyons.

Thus, by measuring the diagonal conductivity, we can detect a Fibonacci anyon. If we apply this procedure for triples
of Fibonacci anyons defining the computational basis (\ref{comp}), by detecting the presence of a Fibonacci anyon in the first pair, 
 we can measure non-destructively the state of the 
Fibonacci qubits.
\section{Relation between the diagonal coset $\PF_k$ and the coset $\widehat{su(2)}_k/ \widehat{u(1)}_{2k}$}
The equivalence of the diagonal coset $\PF_k$ and the $\Z_k$ parafermion realization in terms of the  simpler
coset $\widehat{su(2)}_k/ \widehat{u(1)}_{2k}$, on the level of Virasoro central charges and CFT dimensions of the primary fields, 
is well known \cite{CFT-book,slingerland-bais}. 
In this subsection we will discuss this equivalence from the point of view of the representations of the 
modular group. 

As we have seen in Eq.~(\ref{equiv-S}) the modular $S$ matrix of the diagonal coset (\ref{PF_k}) is identical to that of the affine Lie 
algebra $\widehat{su(2)}_k$.   Here we will show that the diagonal-coset $S$ matrix written in the form (\ref{Sc-2}) also precisely 
coincides with the $S$ matrix for the coset
\beq \label{coset-2}
\frac{\widehat{su(2)}_k}{ \widehat{u(1)}_{2k}}
\eeq
 and will establish an explicit one-to-one correspondence between the primary fields of the two cosets.

First of all we recall the coset construction (\ref{coset-2}) as a CFT model \cite{CFT-book}. The central charge of the numerator is \cite{CFT-book}
\[
c\left(\widehat{su(2)}_k\right)= \frac{3k}{k+2}.
\]
Because the stress-energy tensor of the coset \cite{CFT-book} is defined as the difference of the corresponding tensors of the 
numerator and denominator the central charge of the coset is the difference of the central charges of the numerator and denominator, i.e.
\[
\dot{c}=c\left(\widehat{su(2)}_k\right)-1=\frac{2(k-2)}{k+2}
\]
and coincides with the central charge (\ref{c_PF}). This means that the two realizations of the $\Z_k$ parafermions should be identical 
as CFT models and there should be one-to-one correspondence between primary fields.

The primary conformal fields of the affine Lie algebra $\widehat{su(2)}_k$, which appears in the numerator of the coset~(\ref{coset-2}), 
representing the integrable representations 
are labeled by an integer $l$, such that  $0\leq l\leq k$ as in Eq.~(\ref{L_l}), have CFT dimensions $l(l+2)/[4(k+2)]$ and obey 
$su(2)$ spin-like fusion rules defined in Eq.~(\ref{fusion-su2_k}).

On the other hand, the $\widehat{u(1)}_{2k}$ current algebra in the denominator of the coset~(\ref{coset-2}) is a rational CFT extension of 
the $\widehat{u(1)}$ current algebra with a pair of normal-ordered vertex exponents \cite{kt} 
$U_{\pm\sqrt{2k}}\np{\exp\left( \pm i \sqrt{2k} \phi(z)\right)}$, 
where $\phi(z)$ is a normalized  chiral boson (i.e., with compactification radius $R=1$) \cite{CFT-book}, where $U_{\pm\alpha}$ is an outer 
automorphism of the $\uu$ algebra increasing/decreasing  the $u(1)$ charge by $\alpha$, see Ref.~\cite{kt} for a more precise definition.
This $u(1)$ algebra is understood as the subalgebra of $su(2)$ in the decomposition of an $su(2)$ weight denoted by $l$ into 
$u(1)$ charges like in
\[
(l) \rightarrow \mathop{\bigoplus}\limits_{m=-l}^{l} (m)_1, \quad m\equiv l \mod 2,
\]
where the subscript of $(m)_1$ reminds that these are charges with respect to the $u(1)$ subalgebra, see Eq.~(18.118) in \cite{CFT-book}. 
The primary fields of the $\widehat{u(1)}_{2k}$ current algebra are normal-ordered vertex exponents\cite{kt} with CFT dimensions $\Delta_m$
given below
\beq \label{primary-denom}
U_{m}\np{\ex^{i \frac{m}{\sqrt{2k}} \phi(z)}},  \quad \mathrm{where} \quad  -k+1\leq m\leq k \quad \mathrm{and} \quad \Delta_m =
 \frac{m^2}{4k}.
\eeq
Next, the primary fields of the  coset~(\ref{coset-2})  can naturally  be labeled by two integers $(l,m)$,
first for the numerator and second for the denominator of the coset. The first number $l$ can be interpreted as twice the total $su(2)$ spin, 
while the second one $m$ is twice the spin projection \cite{fat-zam,gep-qiu} so that $-l < m \leq l$.
The CFT dimensions $\Delta^{l}_m$ of the primary fields $\Phi^{l}_m$
are given by \cite{CFT-book}
\[
\Delta^{l}_m = \frac{ l(l+2)}{4(k+2)} -\frac{m^2}{4k}, \quad \mathrm{where}\quad l=0, \ldots, k,\quad -l+1\leq m \leq l 
\]
and $l\equiv m \mod 2$.
It is not difficult to count the number of primary fields for the coset ~(\ref{coset-2}): for each $l$ the number values of $m$  satisfying 
$-l < m \leq l$ is $\#m = (2l+1)-1=2l$ and only half of them will satisfy the parity rule $m\equiv l \mod 2$, so that finally $\#m = l$. Then 
the number of pairs $(l,m)$ satisfying the above conditions is 
\[
\#(l,m)= \sum_{l=0}^k \#m(l) =  \sum_{l=0}^k l =\frac{k(k+1)}{2},
\]
which obviously coincides with the number $ k+1 \choose 2$  of the irreducible representations of the diagonal affine 
coset (\ref{PF_k}) that are labeled by the weights $(\L_\mu+\L_\nu)$ with $0\leq \mu \leq \nu \leq k-1$.

The action of the simple currents on the numerator of the coset~(\ref{coset-2}) induces an action of the simple currents in the denominator
considering the $u(1)$ as a subalgebra of $su(2)$. Because the simple current of $\widehat{su(2)}_k$ is represented by the 
primary field $J=\phi_k$,  corresponding to $l=k$ and has the fusion rules with the primary fields given in Eq.~(\ref{J-su2_k})
the action of the simple current in the denominator should be to map $m \to k- m$. Therefore the simple current's action in the 
coset~(\ref{coset-2}) implies the following field identification  
$
\Phi^{l}_m\equiv \Phi^{k-l}_{k-m}\equiv \Phi^{k-l}_{m-k},
$
where the last identity follows from   $\Phi^{l}_{-m}\equiv \Phi^{l}_{m}$, see Ref.~\cite{CFT-book}. 
On the other hand, because of the obvious identity
$\Phi^{l}_{m+2k}\equiv \Phi^{l}_{m}$ we finally have the field identification in the coset~(\ref{coset-2}) 
\[
\Phi^{l}_m\equiv  \Phi^{k-l}_{m-k} \equiv \Phi^{k-l}_{m+k} ,
\]
depending on which one of $m-k$ and $m+k$ belongs to the range in Eq.~(\ref{primary-denom}).

The modular $S$ matrix for the coset~(\ref{coset-2})  can be written as \cite{CFT-book}
\beqa \label{Sc-3}
S_{(l,m);(l',m')} &=& 2 S^{\widehat{su(2)}_k}_{l,l'} \left( S^{\widehat{u(1)}_{2k}}_{m,m'}\right)^{*}  \nn
&=&  2 \sqrt{\frac{2}{k+2}} \sin\left(  \pi \frac{(l+1)(l'+1)}{k+2}\right) 
\frac{1}{\sqrt{2k}}\ex^{2\pi i \frac{m m'}{2k}}   
\eeqa
so that if we compare this expression with Eq.~(\ref{Sc-2}) we can immediately find the one-to-one correspondence between 
the labels $\L_\mu+\L_\nu$ 
of the diagonal coset (\ref{PF_k}) and the labels $(l,m)$ of the coset~(\ref{coset-2}) as follows
\beq\label{1-1}
\Phi(\L_\mu+\L_\nu) \equiv \Phi^{l}_{m}, \quad l=\nu -\mu, \quad m=\mu+\nu ,
\eeq
where $0\leq \mu \leq \nu \leq k-1$.
More precisely, the comparison of Eq.~(\ref{Sc-3}) with Eq.~(\ref{Sc-2}) implies $l=\nu -\mu \mod 2(k+2)$ and $m=\mu+\nu \mod 2k$, however
due to the above range of $\mu$ and $\nu$ we can  drop the sign ``$ \!\mod$''.
Obviously the branching parity $l \equiv m \mod 2$ is preserved under this identification since the sum and the difference of 
two integers always have the same parity.
\section{Fusion rules in the diagonal coset $\PF_k$ }
\label{sec:fusion}
Now that we have established the one-to-one correspondence (\ref{1-1}) between the primary fields  $\Phi(\L_\mu+\L_\nu)$ 
of the diagonal coset and those of the coset (\ref{coset-2}), derived from the identification of the modular $S$ matrices in both CFT models,
we can use the fusion rules \cite{CFT-book} of the coset  (\ref{coset-2})
\beq \label{fusion-su2_k-coset}
 \Phi^{l}_{m} \times  \Phi^{l'}_{m'} =\mathop{\bigoplus}\limits_{s= |l-l'|}^{\min(l+l', 2k-l-l')} \Phi^{s}_{m+m'},
\eeq
where the summation assumes that $l+l'+s =0 \mod 2$, to derive the fusion rules, given in Eq.~(4.24) in Ref.~\cite{NPB2001},
 in terms of the diagonal coset primary fields $\Phi(\L_\mu+\L_\nu)$.
For this purpose we will need the inverse of the one-to-one correspondence (\ref{1-1})
\beq \label{1-1-inverse}
\mu=\frac{m-l}{2}, \quad \nu=\frac{m+l}{2}, \quad \mathrm{where} \quad m\equiv l \mod 2.
\eeq
Consider first the case when $l+l' \leq 2k-(l+l')$, which is equivalent to $l+l' \leq k$ and implies that $2\min(l,l')\leq l+l'\leq k$,
 hence can be characterized by the condition $\min(l,l')\leq [k/2]$, where $[k/2]$ is the integer part of $k/2$.
In order to take into account the parity rule $l+l'+s =0 \mod 2$ we substitute 
$s= l+l'-2i$ and reverse the order of summation obtaining
\[
\Phi^{l}_{m} \times  \Phi^{l'}_{m'} =\mathop{\bigoplus}\limits_{i= 0 }^{x} \Phi^{l+l'-2i}_{m+m'},
\]
 where $x$ is to be determined from the lower limit $|l-l'| =l+l'-2x$. Assuming $l\leq l'$  (otherwise we can exchange them due to the 
symmetry of the fusion coefficients \cite{CFT-book}) we find $x=l=\min(l,l')$ so that, rewriting the right-hand-side in 
terms of the diagonal coset fields we get
\beq \label{fusion-rules}
\Phi^{l}_{m} \times  \Phi^{l'}_{m'} =\mathop{\bigoplus}\limits_{i= 0 }^{\min(l,l')} \Phi(\L_{\mu''(i)}+\L_{\nu''(i)}) ,
\eeq
where we used the inverse of correspondence (\ref{1-1-inverse}) to write
\[
\mu''(i)=\mu+\mu' +i, \quad \nu''(i)=\nu+\nu' -i. 
\]
To reproduce Eq.~(4.24) in \cite{NPB2001} we set in the above equation $\mu=\mu'=0$ and rewrite Eq.~(\ref{fusion-rules}) as follows
\[
\Phi(\L_{0}+\L_{\nu})  \times  \Phi(\L_{0}+\L_{\nu'}) =\mathop{\bigoplus}\limits_{i= 0 }^{\min(\nu,\nu')} \Phi(\L_{i}+\L_{\nu+\nu'-i}) ,
\]
for $\min(\nu,\nu')\leq [k/2]$,
where we used that according to Eq.~(\ref{1-1}) when $\mu=0$ then $l=\nu$ and when $\mu'=0$ then $l'=\nu'$, so that 
$\min(\nu,\nu') = \min(l,l') = l$ due to the assumption made earlier. 

Next we consider the case when $l+l' > 2k-(l+l')$, which is equivalent to $l+l' > k$ and implies that $2\max(l,l')\geq l+l' >k$,
 hence can be characterized by $\max(l,l') >  [k/2]$. Rewriting Eq.~(\ref{fusion-su2_k-coset}) in this case we have
 \[
 \Phi^{l}_{m} \times  \Phi^{l'}_{m'} =\mathop{\bigoplus}\limits_{s= |\tilde{l}-\tilde{l'}|}^{\tilde{l}+\tilde{l'}} \Phi^{s}_{m+m'}=
 \mathop{\bigoplus}\limits_{i= 0}^{\tilde{x}} \Phi^{\tilde{l}+\tilde{l'} -2i}_{m+m'},
 \]
where we introduced new numbers $\tilde{l}=k-l$, and  $\tilde{l'}=k-l'$, exchanged the sum limits like before and the new lower limit $\tilde{x}$
is obtained from $\tilde{l}+\tilde{l'}-2\tilde{x}=|\tilde{l}-\tilde{l'}|$.
Again we find $\tilde{x}=\min(\tilde{l},\tilde{l'})$ and can rewrite the fusion rule as
\[
\Phi(\L_{0}+\L_{l})  \times  \Phi(\L_{0}+\L_{l'}) 
=\mathop{\bigoplus}\limits_{i= 0 }^{\min(\tilde{l},\tilde{l}')}
 \Phi\left(\L_{\frac{m-\tilde{l}}{2}+\frac{m'-\tilde{l'}}{2}+i}+\L_{\frac{m+\tilde{l}}{2}+\frac{m'+\tilde{l'}}{2}-i}\right) .
\]
Now we can substitute $\tilde{l}$ and $\tilde{l'}$ in terms of $l$ and $l'$ taking into account that $\min(k-l,k-l') = k-\max(l,l')$ to find 
\[
\Phi(\L_{0}+\L_{l})  \times  \Phi(\L_{0}+\L_{l'}) 
=\mathop{\bigoplus}\limits_{i= 0 }^{k-\max(l,l')}
 \Phi\left(\L_{\nu +\nu' -k+i }+\L_{\mu+\mu'+k -i}\right),
\]
where we recall that in the case we consider $\mu=\mu'=0$ so that $l=\nu$ and $l'=\nu'$. Finally we substitute the summation index according to
$j=k-i$ and obtain
\[
\Phi(\L_{0}+\L_{\nu})  \times  \Phi(\L_{0}+\L_{\nu'}) 
=\mathop{\bigoplus}\limits_{j= \max(\nu,\nu') }^{k}
 \Phi\left(\L_{j }+\L_{\nu+\nu' -j}\right), \quad \max(\nu,\nu') >[k/2].
\]
Thus, both cases can be united as follows to reproduce Eq.~(4.24) in Ref.~\cite{NPB2001}
\beq \label{fusion-0-nu}
\Phi(\L_{0}+\L_{\nu})  \times  \Phi(\L_{0}+\L_{\nu'}) = 
\left\{ 
\begin{array}{cc}
\mathop{\bigoplus}\limits_{i= 0}^{\min(\nu,\nu') }
 \Phi(\L_{i }+\L_{\nu+\nu' -i} ) & \mathrm{if} \quad \min(\nu,\nu') \leq [k/2]
\cr 
& \cr
\mathop{\bigoplus}\limits_{i= \max(\nu,\nu') }^{k}
\Phi(\L_{i }+\L_{\nu+\nu' -i}) &  \mathrm{if} \quad \max(\nu,\nu') >[k/2] .
\end{array} 
\right.
\eeq
Finally, using $0\leq \mu \leq \nu \leq k-1$ and  $0\leq \mu' \leq \nu' \leq k-1$, as well as the action of the simple current 
$J=(\L_{1}+\L_{1})$ to represent 
$(\L_{\mu}+\L_{\nu})= J^{\mu}(\L_{0}+\L_{\nu-\mu})$, we obtain the general fusion rules, dropping the symbol $\Phi$ for simplicity,
\beq \label{fusion-mu-nu}
(\L_{\mu}+\L_{\nu})  \times  (\L_{\mu'}+\L_{\nu'}) = 
\left\{ 
\begin{array}{c}
\mathop{\bigoplus}\limits_{i= 0}^{\min(\nu-\mu,\nu'-\mu') }
 (\L_{\mu+\mu'+i }+\L_{\nu+\nu' -i} )  \cr  \mathrm{if} \quad \min(\nu-\mu,\nu'-\mu') \leq [k/2]
\cr \cr \mathrm{or, \ otherwise, } \cr \cr
\mathop{\bigoplus}\limits_{i= \max(\nu-\mu,\nu'-\mu') }^{k}
(\L_{\mu+\mu'+i }+\L_{\nu+\nu' -i})  \cr
  \mathrm{if} \quad \max(\nu-\mu,\nu'-\mu') >[k/2] ,
\end{array} 
\right. 
\eeq
where $[k/2]$ is the integer part of $k/2$. \\

\noindent
\textbf{Example:} \textit{Consider as an illustration the non-Abelian fusion rule between the two primary fields labeled by $(\L_0+\L_2)$ in the 
$\Z_3$ diagonal coset:
i.e., $k=3$ all indices $\mu,\nu, \mu', \nu'$ and $i$ of the weights $\L$ in Eq.~(\ref{fusion-mu-nu}) are defined $\mod 3$,  
while $\nu-\mu=2 >[3/2]$ and $\nu'-\mu'=2 >[3/2]$ so that we have the second case in  Eq.~(\ref{fusion-mu-nu})
\[
(\L_0+\L_2) \times (\L_0+\L_2)=\mathop{\bigoplus}\limits_{i= 2 }^{3} (\L_{i }+\L_{4 -i}) = (\L_{2 }+\L_{2})  \bigoplus (\L_{0}+\L_{1}) ,
\]
which in the notation of Table~3 in Ref.~\cite{NPB2001} is denoted as $\sigma_2 \times \sigma_2 = \sigma_1 \oplus \psi_2$.
}
\section{Full $S$ matrix for the $\Z_k$ parafermion (Read--Rezayi) FQH states}
\label{sec:full}
The full modular $S$-matrix for the $\Z_k$ parafermion FQH states, which determines the modular transformations \cite{CFT-book} of 
the full characters (\ref{full-ch}), has been derived in Ref.~\cite{NPB-PF_k} (see Eq.~(58) there) using  the general properties of the
modular $S$-matrices under the action of simple currents \cite{CFT-book} and the observation that the full character
(\ref{full-ch}) can be represented as a sum over the complete orbit of the repeated action of the simple current of the CFT of
$\uu_{k(k+2)}\otimes \PF_k$
\beq \label{J}
J=\np{\ex^{i\frac{k+2}{\sqrt{k(k+2)}} \phi^{(c)}(z)}} \ \Phi^{\PF}(\L_1+\L_1)(z).
\eeq
It is useful to note that the full character (\ref{full-ch})  can be rewritten \cite{NPB-PF_k}  as a sum over the orbit of the simple current's 
action as
\beq \label{full-ch-2}
\chi_{l,\rho} (\t,\z) = \sum_{s=0}^{k-1} J^s \left[K_{l}(\t,k\z;k(k+2)) \ch(\L_{l-\rho}+\L_{\rho})(\t) \right] .
\eeq
Therefore, one can use here the general property (\ref{S_J}) of the simple currents  \cite{simple-curr} that acting on the $S$-matrix labels they 
only multiply the $S$ matrix by a phase where the $\tilde{Q}_J(\Lambda)$ is the monodromy charge defined in Eq.~(\ref{mon-3})
 \cite{simple-curr,schw}. 

Taking two pairs $(l,\rho)$ and $(l',\rho')$  labeling two representations of the full CFT (\ref{full-CFT}) we can write the full $S$ matrix  
 in terms of the matrix $S^{(1)}$ for the $\uu_{k(k+2)}$ current algebra and the modular matrix $\Sc$ for the diagonal coset as follows 
(cf. Eq.~(58) in Ref.~\cite{NPB-PF_k})
\beq \label{full-S}
S^{l,\rho}{}_{l',\rho'} = k \, S^{(1)}_{l, l'} \ \Sc^{\L_{l-\rho}+\L_{\rho}}{}_{\L_{l'-\rho'}+\L_{\rho'}}
\eeq
where $l, l' $ are defined $\mod k+2$, while $\rho, \rho'$ are $\mod k$ indices, the $\uu_{k(k+2)}$ modular matrix  $S^{(1)}_{l, l'}$ 
is given by \cite{NPB-PF_k}
\beq \label{S^1}
S^{(1)}_{ll'}  =\frac{1}{\sqrt{k(k+2)}} \ex^{-2\pi i\frac{l l'}{k(k+2)}}
\eeq
while $\Sc^{\L_{l-\rho}+\L_{\rho}}{}_{\L_{l'-\rho'}+\L_{\rho'}}$ is the $S$ matrix  for the diagonal coset (\ref{PF_k})
defined in Eq.~(\ref{S-coset}).

Finally, we can use the compact formula (\ref{Sc-2}) for the coset $S$ matrix and the explicit formula (\ref{S^1}) for the $\uu_{k(k+2)}$
to write the full modular  $S$ matrix for the $\Z_k$ parafermion FQH states
in the basis of the full characters $\chi_{l, \rho}$ defined in Eq.~(\ref{full-ch})
\beqa \label{full-S-comp}
S^{l,\rho}{}_{l',\rho'} &=& \frac{2 }{k+2}  \exp\left(2\pi i \frac{l l'}{k+2}\right) \times \nn
 && \sin\left( \frac{\pi \left([2\rho - l ]_{\! \mod k} +1\right) \ ( [2\rho'-l']_{\! \mod k} +1) }{k+2}\right) ,
 \eeqa
 where the indices $l, l'$ are defined $\mod (k+2)$, $\rho, \rho'$ are defined $\mod k$ and $l-\rho \leq \rho \mod k$ and 
$l'-\rho' \leq \rho' \mod k$.
Another compact form of the $S$-matrix, however only  for the $\Z_3$ parafermion FQH state, can be found in Eq.~(5.7) in Ref.~\cite{NPB2001}.
\section*{Conclusions}
Using the level--rank duality for affine Lie algebras  we derived in this paper a compact analytic expression for the full modular $S$ matrix 
of the $\Z_k$  parafermion FQH state proposed by Read and Rezayi. This matrix contains a charged component, coming from the $\uu$ 
sector, representing the electric charge of the quasiparticles and the Aharonov--Bohm effect, and a neutral component corresponding 
to the $\Z_k$  parafermions realized as a diagonal affine coset model.
This compact expression can be used to calculate the interference patterns of non-Abelian anyons and compare them to the 
experimental results in order to eventually identify them in real physical systems. We have also derived rigorously the fusion rules in the 
diagonal affine coset model using its relation with the $su(2)_k/u(1)$ coset model.
\ack
I thank Andrea Cappelli and Ady Stern for many useful discussions as well as  the Galileo Galilei 
Institute for Theoretical Physics in Firenze, Italy for hospitality. The author has been supported as a Research Fellow by 
the Alexander von Humboldt Foundation and by the Bulgarian Scientific Fund under Contract No. DN 18/3 (2017) .\\
\appendix
\section{$S$ matrix for $k=3$}
\label{app}
In this appendix we compute directly the modular $S$ matrix for $\widehat{su(3)}_2$ using Eq.~(\ref{S-duality}).
For general $k$ the $\widehat{su(k)}_2$ vacuum--vacuum element of the $S$ matrix corresponds 
to $\l=\L_0+\L_0$, $\l'=\L_0+\L_0$ so that $|\l|=|\l'|=0$, hence ${}^t\l={}^t\l' = 0$ and
\beq\label{S-00}
S^{\widehat{su(k)}_2}_{\L_0+\L_0, \L_0+\L_0} = \sqrt{\frac{2}{k}} \ 
S^{\widehat{su(2)}_k}_{0,0} = \frac{2}{\sqrt{k(k+2)}}\sin\left(\frac{\pi}{k+2}\right).
\eeq
For the $k=3$ case we will express the $S$ matrix in terms of the Golden ratio $\delta$ defined by
\beq\label{delta}
\frac{\delta}{2} \equiv \cos\left( \frac{\pi}{5}\right) \quad \Longleftrightarrow \quad
\delta^2=\delta+1 \quad \left( \Rightarrow \delta=\frac{1+\sqrt{5}}{2}\right) ,
\eeq
where we have taken the positive root of the quadratic equation. Next we express  $\sin(\pi/5)$ in terms of $\cos(\pi/5)$
using 
\[
\sin\left( \frac{\pi}{5}\right)=\sqrt{1- \cos^2\left( \frac{\pi}{5}\right)} =\sqrt{1-\left( \frac{\delta}{2}\right)^2}
=\frac{\sqrt{3-\delta}}{2}
\]
where we exploited the quadratic equation in  Eq.~(\ref{delta}). It is a general property of the modular $S$ matrix that its 
vacuum--vacuum element is equal to the inverse of the \index{total quantum dimension} total quantum dimension $\Dc$, i.e.,
\beq\label{S-00.2}
S^{\widehat{su(k)}_2}_{\L_0+\L_0, \L_0+\L_0}\equiv \frac{1}{\Dc}, \quad
{\Dc}^2 \mathop{=}^{\mathrm{def}} \sum_i d^2_i,
\eeq
where the sum over $i$ goes over all integrable irreducible representations. As we will see later, among the 
$6$ integrable representations of $\widehat{su(3)}_2$ three have quantum dimension $1$ and three have quantum dimension $\delta$.
Therefore, in this case we must have
\beq \label{Dc}
\Dc^2=3. 1^2 +3 . \delta^2=3(1+\delta^2)=3(\delta+2) \ \
\Rightarrow \ \ \Dc=\sqrt{3(\delta+2)},
\eeq
where we used again Eq.~(\ref{delta}). Below we will show that $1/\Dc=(2/\sqrt{15})(\sqrt{3-\delta})/2$. Indeed,
\[
\frac{3-\delta}{15} -\frac{1}{\Dc^2} = \frac{1}{3}\left( \frac{3-\delta}{5} - \frac{1}{\delta+2}\right)=
\frac{1}{3}\frac{(3-\delta)(\delta+2)-5}{5(\delta+2)} \equiv 0,
\]
exploiting  Eq.~(\ref{delta}) once again, taking into account that if $a^2-b^2=0$ and $a+b>0$ then $a=b$.
Thus we prove that
\beq\label{S-00.3}
S^{\widehat{su(3)}_2}_{\L_0+\L_0, \L_0+\L_0} = \frac{2}{\sqrt{15}}\sin\left(\frac{\pi}{5}\right)=
\sqrt{\frac{3-\delta}{15}}= \frac{1}{\Dc}.
\eeq
In Table~\ref{tab:charges} we show the monodromy charges (\ref{mon-2.2}) for the integrable representations 
of $\widehat{su(3)}_2$, labeled  by $\L_\mu+\L_\nu$, $0\leq \mu \leq \nu \leq 2$, with respect to the simple current 
represented by the weight $J\equiv \L_1+\L_1$.
\begin{table}[htb]
\caption{Monodromy charges of integrable representation for $\widehat{su(3)}_2$. \label{tab:charges}} 
\centering
\begin{tabular}{|c||c|c|c|c|c|c|}
\hline
 $\L_\mu+\L_\nu$ & $\L_0+\L_0$ &  $\L_0+\L_1$ &$\L_0+\L_2$ & $\L_1+\L_1$ & $\L_1+\L_2$ & $\L_2+\L_2$ \cr\hline\hline
 & & & & & & \cr
 $\tilde{Q}_J(\L_\mu+\L_\nu)$ & 0 &  $-\frac{1}{3} $ & $ -\frac{2}{3} $ & $-\frac{2}{3}$ & $- 1$ & $-\frac{1}{3}$ \cr
 & & & & & $(0 \mod \Z) $ & \cr
\hline
\end{tabular}
\end{table}
These monodromy charges will be used  for the computation of the elements of the modular $S$ matrix
using the property (\ref{S_J}).
Now we can obtain the other elements in the first row of the $S$ matrix using Eq.~(\ref{S_J}) and Table~\ref{tab:charges}
\beqa
S^{\widehat{su(3)}_2}_{\L_0+\L_0, \L_1+\L_1} &=&S^{\widehat{su(3)}_2}_{\L_0+\L_0, J*(\L_0+\L_0)}= \ex^{-2\pi i \tilde{Q}_J(\L_0+\L_0)}
S^{\widehat{su(3)}_2}_{\L_0+\L_0, \L_0+\L_0} \nn 
&=&\frac{1}{\Dc}, \nonumber
\eeqa
since $\tilde{Q}_J(\L_0+\L_0)=0$. Similarly $S^{\widehat{su(3)}_2}_{\L_0+\L_0, \L_2+\L_2}=1/\Dc$.
In order to find the other three matrix elements we have to use Eq.~(\ref{S-duality}) directly: 
\[
S^{\widehat{su(3)}_2}_{\L_0+\L_0, \L_0+\L_1}=\sqrt{\frac{2}{k}} \ex^{2\pi i \frac{0|\L_0+\L_1|}{2k}}S^{\widehat{su(2)}_3}_{0,1}=
\frac{2}{\sqrt{k(k+2)}}\sin\left( \frac{2\pi}{k+2}\right),
\]
where $ \l'=\L_0+\L_1$ so that  $ {}^t\l'=(\L_0+\L_1)^t= {\tiny \yng(1)}^t={\tiny \yng(1)}$ and therefore we used $l=0$, $l'=1$.
For $k=3$ we have
\[
\frac{2}{\sqrt{15}}\sin\left( \frac{2\pi}{5}\right)=\frac{2}{\sqrt{15}}2\sin\left( \frac{\pi}{5}\right)\cos\left( \frac{\pi}{5}\right) =\frac{\delta}{\Dc},
\]
where we used Eqs.~(\ref{S-00.3}) and (\ref{delta}). Similarly, because  $\tilde{Q}_J(\L_0+\L_0)=0$, we find
$S^{\widehat{su(3)}_2}_{\L_0+\L_0, \L_1+\L_2}=\delta/\Dc$ and so on.
Next,  taking $\tilde{Q}_J(\L_1+\L_1)=-2/3$ from Table~\ref{tab:charges}, we compute the diagonal element
\beqa
S^{\widehat{su(3)}_2}_{\L_1+\L_1, \L_1+\L_1}&=&S^{\widehat{su(3)}_2}_{J*(\L_0+\L_0), \L_1+\L_1}= \ex^{-2\pi i \tilde{Q}_J(\L_1+\L_1)}
S^{\widehat{su(3)}_2}_{\L_0+\L_0, \L_1+\L_1} \nn &=&\frac{\ex^{-\frac{2\pi i}{3}}}{\Dc}. \nonumber
\eeqa
Another important diagonal element is 
\beqa
S^{\widehat{su(3)}_2}_{\L_0+\L_1, \L_0+\L_1}&=&\sqrt{\frac{2}{k}} \ex^{2\pi i \frac{|\L_0+\L_1|^2}{2k}}S^{\widehat{su(2)}_k}_{1,1} \nn &=&
\frac{2}{\sqrt{k(k+2)}} \ex^{i \frac{\pi}{k}}\sin\left( \frac{4\pi}{k+2}\right)=\frac{\ex^{i\frac{\pi }{3}}}{\Dc}, \nonumber
\eeqa
where we used $k=3$ and $\sin(4\pi/5)=\sin(\pi/5)$.
Next, taking $\tilde{Q}_J(\L_0+\L_1)=-1/3$ from Table~\ref{tab:charges}, we compute the element
\beqa
S^{\widehat{su(3)}_2}_{\L_0+\L_1, \L_1+\L_2}&=&S^{\widehat{su(3)}_2}_{\L_0+\L_1, J*(\L_0+\L_1)}  = \ex^{-2\pi i \tilde{Q}_J(\L_0+\L_1)}
S^{\widehat{su(3)}_2}_{\L_0+\L_1, \L_0+\L_1}   \nn &=&\ex^{-2\pi i\frac{(-1)}{3}}\frac{\ex^{i\frac{\pi }{3}}}{\Dc}=\frac{-1}{\Dc}. \nonumber
\eeqa
As a final illustration, taking  $\tilde{Q}_J(\L_1+\L_2)=0 \mod \Z$ from Table~\ref{tab:charges}, we compute the element
\beqa
S^{\widehat{su(3)}_2}_{\L_1+\L_2, \L_1+\L_2} &=&
S^{\widehat{su(3)}_2}_{J*(\L_0+\L_1), \L_1+\L_2}= \ex^{-2\pi i \tilde{Q}_J(\L_1+\L_2)}
S^{\widehat{su(3)}_2}_{\L_0+\L_1, \L_1+\L_2} \nn &=&\frac{-1}{\Dc}. \nonumber
\eeqa

\section*{References}
\bibliography{CB,FQHE,Z_k,my,TQC}

\end{document}